\documentclass{article}
\usepackage{aas2pp4}

\def\alwaysmath#1{\ifmmode{#1}\else{$#1$}\fi}

\def\teff{\alwaysmath{T_{\rm eff}}}
\def\kelvin{{\rm \, K}}
\def\kms{\alwaysmath{{{\rm \,km\,sec^{-1}}}}}
\newcommand\hst{{\it HST}}
\newcommand\ngc[1]{NGC\,#1}
\newcommand\etal{et~al.}

\slugcomment{Submitted to ApJ Letters}

\lefthead{Ferraro et al.}
\righthead{HST Photometry of M3 \& M13}

\begin{document}

 \title{HST UV Observations of the Cores of M3 and M13\footnote{Based on
observations with the NASA/ESA {\it Hubble Space Telescope}, obtained at
the Space Telescope Science Institute, which is operated by AURA, Inc.,
under NASA contract NAS5-26555}}

\author{Francesco R. Ferraro\altaffilmark{2},  
Barbara Paltrinieri\altaffilmark{2},  
Flavio Fusi Pecci\altaffilmark{2,3},  
Carla Cacciari\altaffilmark{2,4}},
\author{Ben Dorman\altaffilmark{5,6},
Robert T. Rood\altaffilmark{5}}

\altaffiltext{2}{Osservatorio Astronomico di Bologna, via Zamboni 33, I-40126
Bologna, ITALY}
\altaffiltext{3}{Stazione Astronomica di Cagliari, 09012 Capoterra, ITALY}
\altaffiltext{4}{Space Telescope Science Institute, Baltimore, USA}
\altaffiltext{5}{Astronomy Dept, University of Virginia,
	P.O.Box 3818, Charlottesville, VA 22903-0818}
\altaffiltext{6}{Laboratory for Astronomy \& Solar Physics, 
Code 681, NASA/GSFC,	Greenbelt MD 20771}

\begin{abstract}

We present preliminary results from HST/WFPC2 observations of the
central regions of the of the Galactic globular clusters M13 \& M3.
The clusters are almost identical in most respects including chemical
composition, but there are dramatic differences in both the
horizontal branch (HB) and blue straggler (BSS) populations. The M13 HB
has a long blue tail extending 4.5 mag in $V,$ reaching well below the
level of the main sequence turn-off. M3 has no such feature.
M3 \& M13 are thus an extreme case of the ``second parameter problem''
in HB morphology.  Also present in the M13 HB are two gaps similar to
those seen in the clusters NGC 6752 and NGC 2808.  M3 has a specific
frequency of BSS three times larger than that of M13.  Our results
imply that neither age nor cluster density, two popular second
parameter candidates, are likely to be responsible for the observed
differences.

\end{abstract}


\section{Introduction}
We are involved in two projects obtaining Hubble Space Telescope (\hst)
photometry of globular cluster stars. The goal of the first is to
obtain the largest most complete possible samples of a number of
prototype clusters. The second is an investigation of clusters
suspected or known to have a significant population of UV bright
stars.  Here, we report preliminary results for observations of the
central regions of one cluster from each project, M3 (\ngc{5272}) from
the first and M13 (\ngc{6205}) from the second.

Both clusters are nearby, northern hemisphere clusters with low
reddening (see Table 1). Their metallicities are very 
similar---indeed in many respects the clusters are 
almost twins. M3 has
been well studied photometrically dating back to \cite{san53}. Its
color-magnitude diagram (CMD) often serves as the text book example
used to demonstrate globular cluster sequences. M13 was also a target
of early photometric studies (\cite{arp55}). From the outset its CMD, in
particular the horizontal branch (HB), was obviously different from
that of M3. After the emergence of the ``second parameter'' problem
(e.g., \cite{sw67}; \cite{rood73}), M3 \& M13 could have been
discussed as a second parameter pair, although they have not been
discussed in that context as often as, e.g., the \ngc{288}/\ngc{362} pair
(with the notable exception of Catelan \& de Freitas Pacheco 1995).
M13 was chosen as one of our \hst\ targets because of its very blue
far-UV color.  However, our results are so striking in the context of
the second parameter problem that we present preliminary results here.

\section{Observations}

We have obtained WFPC2 observations of the galactic globular clusters M3
(Cycle 4: GO-5496; PI F. Fusi Pecci) and M13 (Cycle 5: GO-5903; PI F.R.
Ferraro). We describe here results obtained using the $V$ (F555W), $U$
(F336W) and the mid-UV filters (F255W) mapping the cluster cores.  The
CMDs presented here are {\it preliminary results} of the four WFPC2 fields
obtained with the the PC located on the cluster center.  The exposures
taken in each filter 
in M3 are listed in Table~1 by \cite{m3bss97}. The total times in M 13
are 32s in F555W, 560s in F336W and 200s in F225W. 
Details of the data reduction will be given elsewhere. 
A brief description of the procedure can be found in 
\cite{m3bss97}.

\section{Results}

\begin{figure}
\plotone{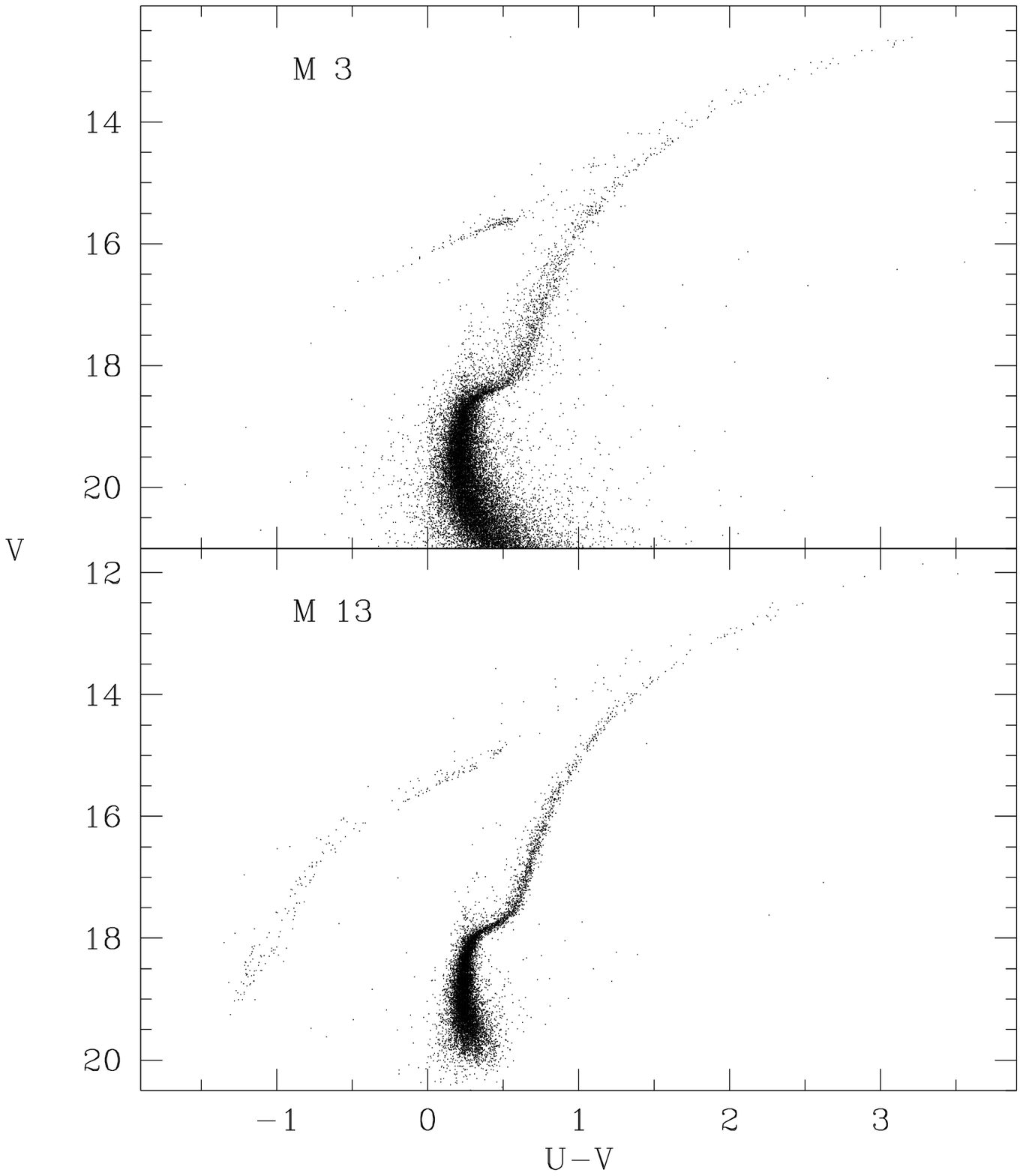}
\caption{
\label{vuvcmd}
The $(V,U-V)$ color magnitude diagram for M3 (upper) and M13 (bottom). 
Measurements
were taken in the HST F336W and F555W filters and transformed to the Johnson
system using the calibrations given in \protect\cite{holt1995}. 
}
\end{figure}

Figure~\ref{vuvcmd} shows the $V,~U-V$ CMDs for the two clusters. 
All stars measured in the HST field (PC1 $+$ WFCs) are shown. The M3
sample contains more than 29,000 stars; 
in M13 there are $\sim  11,000$ stars, and the limiting
magnitude is about 0.5 mag brighter. The main sequences are clearly defined
from the red giant branch (RGB) tip down to about 2 mag below the main
sequence turnoff (MS-TO). 

The MS, sub-giant branch (SGB), and the red giant branch (RGB)
are quite similar in the two clusters. The asymptotic giant branch (AGB)
of M3 is somewhat more prominent than in M13. However, the HBs and blue
straggler sequences are dramatically different.

The M13 HB extends much further to the blue than that of M3.  There is
a very long HB blue tail, extending about 1.5 mag below the MS-TO.  The HB
shows two gaps at $V\sim 15.9$ and $V\sim 17.8$. Except for a few stars the M3
HB terminates at a point corresponding to the upper gap in M13.  

We defer discussion of the statistical significance of the gaps until
the final data reduction has been performed. Since other clusters
(\ngc{6752}---\cite{buo86}; \ngc{2808}---\cite{ferraro2808} and
\cite{sos97}) show similar gaps, we adopt their reality as a working
hypothesis.  M13's long blue HB tail has not been seen in earlier
published CMDs (\cite{gua93};
\cite{mj94}) which did not reach sufficiently deep. We had anticipated
its existence on the basis of the very blue integrated $(15-V)$ color
(\cite{vdd81}; \cite{dor95}).  The blue tail is present in two
unpublished ground based studies of the outer parts of the cluster.
Stetson (1996) gives a preliminary CMD based on CFHT
$B,~I$ photometry which shows an HB extending well below the MS-TO,
similar to ours. The lower HB gap is probably present in Stetson's
data.  The upper gap is not present, but it is difficult to evaluate
this difference because the different photometric systems and
preliminary state of both data sets.  In another preliminary reduction,
\cite{m13bv97} present $B,~V$ data obtained with the Calar-Alto 1.23\,m
telescope for stars $> 200\arcsec$ from the cluster center. The HB
reveals the long blue tail and shows both of the gaps present in the
HST data.

\begin{figure}
\plotone{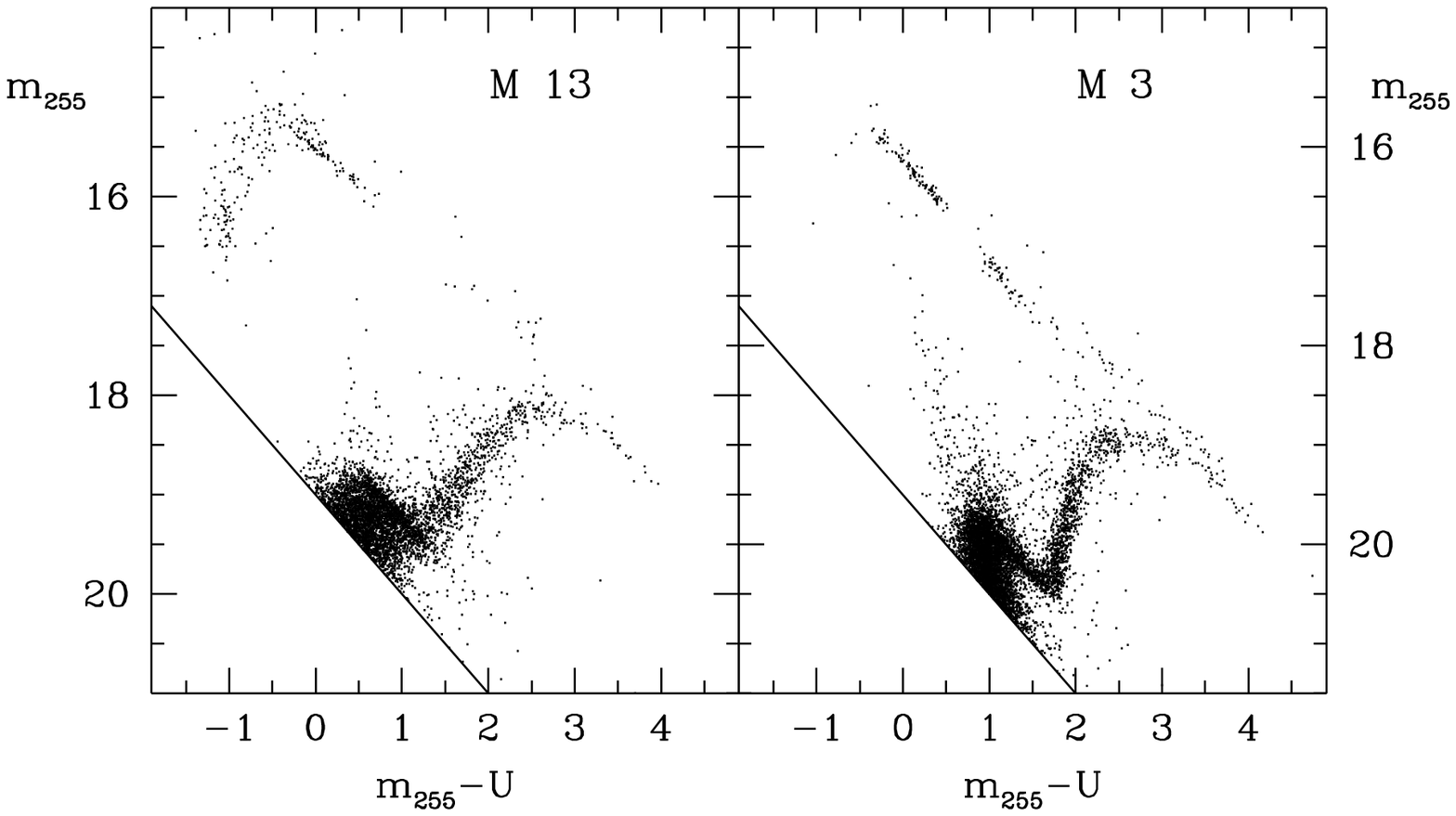} 
\caption{
\label{uuucmd}
The $(U,~m_{255}-U)$ color magnitude diagram for M13 (left) and M3 (right). Measurements
were taken in the HST F255W and F336W filters. The F336W data has been transformed
to the Johnson system as in Fig.~\protect\ref{vuvcmd}. 
The $m_{255}$ magnitudes are on the STMAG system.}
\end{figure}

Note that the traditional RR~Lyrae ``gap'' occurs at the right end of
the M13 HB. The $U-V$ color is not very sensitive to \teff\ in the
range of the RR~Lyrae and red HB. The entire ``horizontal'' part of the
M3 HB (i.e., $V\approx {\rm constant}$) is compressed into such a small
range that no part of the HB appears horizontal in the $V,~U-V$ CMD,
and the cool end of the HB terminates in a clump. 
The hotter part of the M3 HB ($U-V<0.3$) overlaps the corresponding
portion of the M13 HB.  The
redder part of the HB can be seen more easily in the
$m_{255},~m_{255}-U$ CMD (Fig.~\ref{uuucmd}).  The location and
morphology of the main branches seen in these colors is strikingly
different from that produced by optical filters:
 the normally bright RGB is actually faint here, and the HB is far from being
 ``horizontal''.  The M3 HB is seen as a very narrow, straight
``diagonal'' branch extending upward to the left away from the giant
branch. The obvious ``gap'' is the RR~Lyrae gap. The M3 HB terminates
just where the HB begins to turn down  $(\teff
\gtrsim 12000 \kelvin)$
in $m_{255}$ because of the increasing bolometric correction. The M13
HB starts on the hot side of the RR~Lyrae gap, rises to the maximum
where significant bolometric corrections set in, and the long blue HB
tail is compressed into a very short sequence. For stars cooler than
$U-V \sim 0.3,$ the M13 stars are more luminous than those in M3 in
the comparable color range.  This result is expected if the reddest HB
stars in M13 are near or past core helium exhaustion and on their way
to the lower AGB.

In Fig.~\ref{vuvcmd} M13 shows a 
small population of Blue Straggler Candidates (BSS),
about a dozen, which are clearly separated from the main branches. The
population of BSS in M3 is far larger. Indeed there are so many faint BSS
in M3 that it is easy to mistake them for a distortion of the SGB near the
turnoff.  One can more cleanly select BSS in the ($m_{255},~m_{255}-U$) CMD
(Fig. 2). The solid horizontal line at $m_{255}=19$ is the magnitude limit
used by \cite{m3bss97} in their detailed study of the M3 BSS to separate
{\it bright} from {\it faint} BSS.  Again we see many more BSS in M3 than
M13.  Specifically, in M3, \cite{m3bss97} find 171 BSS stars: 72 in the
bright sample and 99 in the faint one.  M13 has only 35 BSS stars: 11 in
the bright sample and 24 in the faint one.  

 Note that the light sampled in the two clusters
turns to be almost the same:
$\sim 10 \times 10^4 L_{\odot}$ 
and $\sim 12 \times 10^4 L_{\odot}$ 
 for M13 and M3 respectively, so that the number of BSS per 
unit of sampled luminosity in M3 turns to be
about three times higher than in M13.

\section{Discussion}

\begin{figure}
\plotone{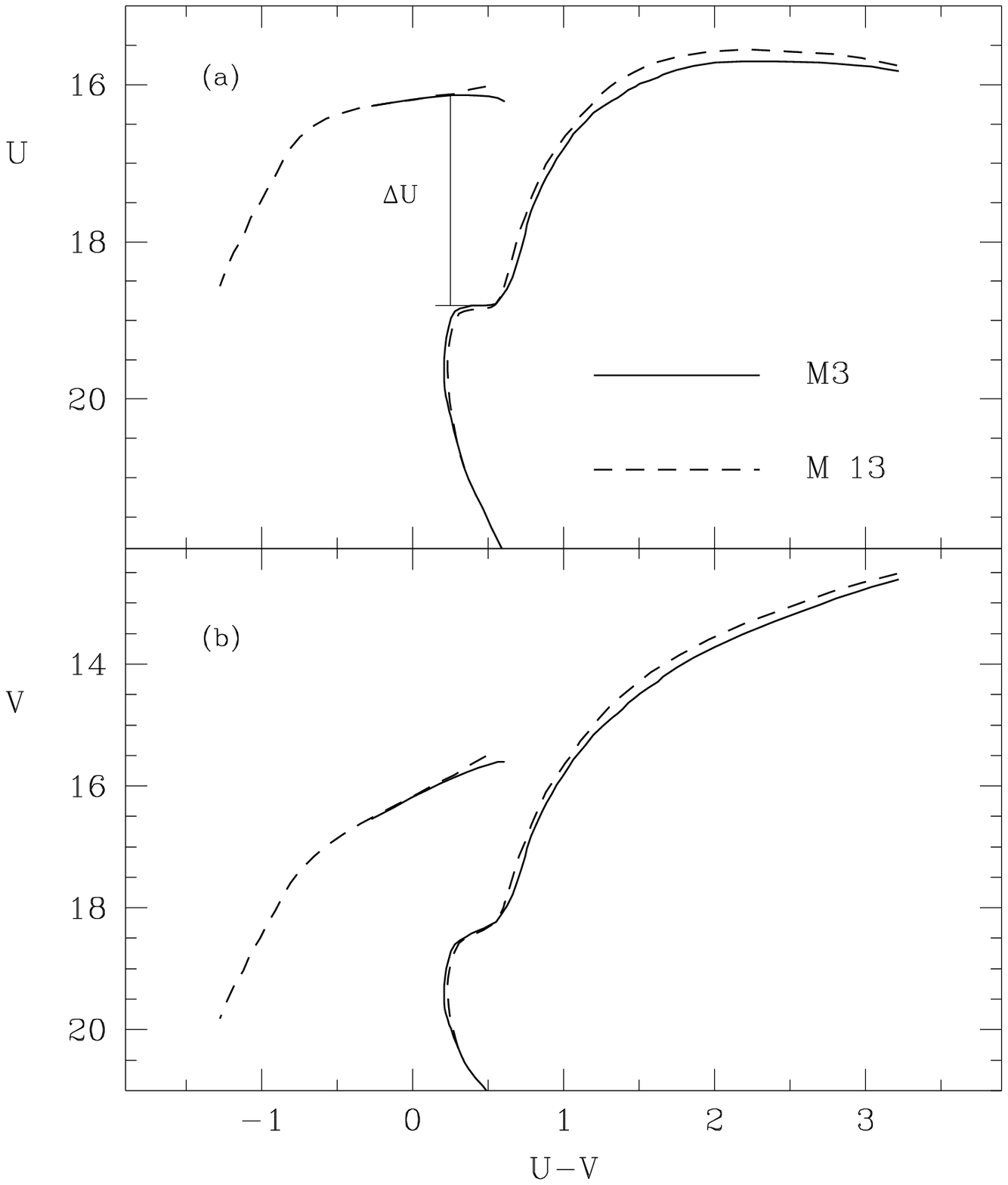}
\caption{
\label{ridge}
Ridge lines for the principal sequences in M3 (solid line) and M13 
(dashed line) superposed by a shift
of 0.6 mag. in the 
$U,~U-V$ {\it panel (a)} and $V,~U-V$ {\it panel (b)}
CMDs.
The observed difference between the HB and SGB ($\Delta U$) is indicated.
}
\end{figure}

Table~1 summarizes many of the parameters of the two clusters.  Given the
almost identical [Fe/H], we can discuss the differences in HB
morphology in terms of the second parameter problem. The second
parameter most often considered is age (e.g., \cite{ldz94}).  
To check for an age difference between M13 \& M3 we have
drawn the mean ridge through the main branches, adopting the usual
procedure which eliminates the most discrepant objects.  Figure~\ref{ridge}
shows the result obtained by overlapping the cluster mean loci
aligning the HBs where the ZAHB is well populated in both clusters.
This was accomplished by applying a vertical shift 0.6 mag to the M13
data. We plot both the $V,~U-V$ and $U,~U-V$ CMDs.
The MS,
SGB, and lower RGB mean ridge lines are seen to be almost coincident. 
(We do not consider
the small differences shown in the plot to be significant.) 
The M13 upper RGB is sparsely populated and it is not clear at this
point whether the difference in ridge lines shown in Fig.~\ref{ridge} is
significant. 

\begin{deluxetable}{lccl}
\tablewidth{\columnwidth}
\tablecaption{Basic Information on the Target Clusters}
\tablehead{
 \colhead{}           & \colhead{M3}      &
\colhead{M13} & \colhead{Ref.}
}
\startdata
 $(m-M)_V$ & 15.05 & 14.35 &  (1) \nl
 $E(B-V)$ & 0.01 & 0.02 &  (1) \nl
 [Fe/H] & $-1.47 \pm 0.01$ & $-1.51 \pm 0.01$ & (2) \nl
 $(15-V)_0$ & 3.41 & 1.63 & (3) \nl
$c$ & 1.85 & 1.5 & (4) \nl
 $\log{\rho}_0$ & 3.5 & 3.4 & (4) \nl
 $\log (M/M_{\odot})$ & 5.8 & 5.8 & (4) \nl
 $\epsilon$ & 0.04 & 0.11 & (5) \nl
\enddata
\tablenotetext{(1)}{\protect\cite{peterson93}}
\tablenotetext{(2)}{\protect\cite{kraftm13m3}}
\tablenotetext{(3)}{\protect\cite{dor95}}
\tablenotetext{(4)}{\protect\cite{trager93}}
\tablenotetext{(5)}{\protect\cite{ws87} }
\end{deluxetable}

 The mean metallicities and the CNO abundances (${\rm C + N + O}$) are
similar (Table 1), and there is no indication of any significant
peculiarity in the primordial helium abundance. Thus, the clusters must
have essentially the same age; if not the sequences would not match over
such an extensive range in the diagram. We can estimate the maximum age
difference best from the $U,~U-V$ CMD.  The most age
sensitive feature in this CMD is the difference between the HB and the
SGB ($\Delta U$, see Fig.~\ref{ridge}). 
From the isochrones of Dorman (1996, unpublished) we estimate that
the level of the almost horizontal SGB decreases by about 0.06\,mag/Gyr.
Our observed difference between M13 and M3 $\delta \, \Delta U$ is
0.06--0.1\,mag.  On this basis we estimate that M13 is $\la 1$--1.5\,Gyr
older than M3.  The precision of our result is somewhat compromised by the
low sensitivity of the $U-V$ color to \teff\  near turn-off. 
On the other hand, the data for the two clusters were obtained by the
same instrument using the same filters.  Early studies of cluster CMDs
were often hampered by limited dynamic range, whereby the bright and faint
sequences were exposed at different times with different equipment. Our
observations suffer none of these problems.  Thus we avoid at least some
of the possible pitfalls which can undermine differential age analyses
(see the review in 
\cite{svb96}). 

Earlier studies which specifically addressed the differential
age of the M13/M3 pair have achieved similar results.
\cite{vbs90} used the ``horizontal'' method, in which age is derived from
the color difference between the turnoff and the lower RGB.
Unfortunately, the CMDs available to them that also had well-defined
turnoffs had very sparsely populated RGB's. They estimated that M13
was 0.9--2\,Gyr older than M3. The ``vertical'' or $\Delta
V_{TO}^{HB}$ method, using the magnitude difference between the
turnoff and RR~Lyrae to determine age, has suffered from the fact that
M13 has few RR~Lyrae which might well be evolved and thus more
luminous than the M3 RR~Lyrae.  Our results confirm that the reddest
HB stars in M13 are significantly evolved away from the ZAHB.
\cite{catelanm3m13} have used theoretical HB models to correct for
this and offer a very detailed discussion of relative age of the pair.
They conclude that there is essentially no age difference with an
error of $\pm3$\,Gyr.  Chaboyer, Demarque \& Sarajedini (1996) estimate,
based on the $\Delta V_{TO}^{HB}$ technique, that M13 is $\approx 2\,$Gyr 
older than M3. We conclude in concordance with \cite{svb96},
that 2 Gyr is an upper bound to the age difference.

Could age be responsible for the dramatically different HB
morphologies?  The work of \cite{catelanm3m13} implies otherwise, and
our data adds force to this conclusion. The extended tail we have
observed in M13 implies the presence of very low mass HB stars.  To
produce such stars by ``aging'' M3 would require a large change in
turnoff mass (and thus age).  The required age difference would be
even larger than the $\sim 5\,$Gyr found by \cite{catelanm3m13} from
synthetic HB models.  Whitney \etal\ (1997) using HB Hess diagrams
obtain a similar estimate. In addition, they find that the mass
dispersion would have to be more than twice as large in M13 as in M3.
Simple measures of HB morphology such as that of Lee \etal\
(1994), which measure the ratio of stars relative to the RR Lyrae
strip, can change dramatically within 2 Gyr. These parameters,
however, do not adequately represent the differences in the HBs of M13
as compared to M3.  Since an age difference 
large as the $\sim 5\,$Gyr required to model the observed HB
distributions seems to be ruled out, we conclude that {\em age cannot be the
driving second parameter}, at least not for this classic pair.

\cite{fusipecciBT} have shown that some second parameter
characteristics, especially long blue HB tails, are correlated with
cluster structural parameters like central density. Yet M3 \& M13 have
structural parameters that are identical within the errors (Table~1).
So again for this pair, the second parameter is not associated with
central density or concentration.  

The similarity of the structural parameters raises questions about the
origin of the BSS. BSS are thought to originate from stellar
collisions and/or merging primordial binaries. \cite{m3bss97} argue
that the radial distribution and luminosity functions of the M3 BSS
are consistent with a collisional origin for the central BSS and a
merged-binary origin for those in the outer cluster.  A paucity of BSS
in the exterior of M13 could arise either because there were fewer
binaries to begin with or because the binaries were destroyed.
However, the low specific frequency of central BSS in M13 as compared
to M3 is very puzzling.  Perhaps, the mechanism producing BSS in the
central region of M3 is more efficient than in M13 because M3 and M13
are experiencing different dynamical evolutionary phases.
Alternatively, \cite{bp95} have suggested that M3 is in a short
lived phase of its dynamical evolution during which its population of
primordial binaries in being rapidly coverted to BSS.

It is also interesting to note that blue tails do not seem to
correlate with BSS either in position within a given cluster or from
cluster to cluster (\cite{sos97}). This suggests that BT's do not
arise from binary stars or stellar interactions.

What are known differences between M3 \& M13?  

Peterson, Rood, \& Crocker (1995) showed that perhaps 1/3 or M13's HB
stars had relatively rapid rotation, whereas no rapid rotators were found
in M3. Also supporting the idea that M13 stars on the average rotate more
rapidly than those in M3 are the observations of \cite{kraftm13m3} and
\cite{lickm13m396}. They find evidence for substantially more processing
of CNO elements in the envelopes of M13 RGB stars than those in M3. This
mixing of processed material to the surface could be plausibly linked to
rotation. Stellar rotation has long been suggested as a mechanism to
enhance mass loss (Renzini 1977, Fusi Pecci 
\& Renzini 1978, Peterson \etal\ 1995) which would produce a bluer HB. 
It is not obvious how it would relate to BSS production efficiency.
M13 is more elliptical than M3.  \cite{norris83} noted a possible
correlation between HB-morphoplogy, rotation and ellipticity.  In the
case of M13, \cite{m13rvel} show that the ellipticity is related to
the overall rotation of the cluster at a rate about 5\kms. It is not
clear that cluster rotation would affect rotation of individual stars
or that rotation rates set at the time of stellar formation would ``be
remembered'' in later stages of evolution (e.g., \cite{pksd89}). We
mention rotation and ellipticity only because they are among the few
parameters which have been observed to differ both in
the mean between the clusters and on a star-to-star basis.

\cite{catelanm3m13} also noted the difference in RGB CNO processing in
their study of M3 \& M13. They searched for similar correlations in other
clusters. In particular they note that the most highly processed
super-oxygen-poor stars (SOP) of \cite{kraftm13m3} are found only in
clusters with very blue HBs. \cite{sweigart97} has outlined a scheme which
would directly connect the SOP phenomenon to bluer HBs. He has pointed out
that the higher Na and Al over-abundances in the SOP stars could only
arise if helium was being mixed to the surface. The following HB phase of
such stars would be bluer and slighly more luminous than for non-He
enhanced stars. While He mixing may play a role in the M13/M3 difference we
feel it  cannot be the whole story. Suppose one considered ``processing'' an HB
by turning on rotation which would lead to mixing in some stars. The
non-rotating stars would be unaffected. Basically the red end of the HB
would remain relatively unchanged while a blue tail grows.  
M3 has a significant population of red HB
stars and RR~Lyrae; M13 does not. To ``convert'' M3 to M13, all stars
would have to undergo helium enhancement. Observations suggest otherwise.
\cite{kraftm13m3} and \cite{lickm13m396} find that not all M13 giants are
SOP; some have relatively normal O. Some stars reach the HB without
significant O processing---\cite{prc95} find that O in the redder HB
stars of M13 is more-or-less normal. Indeed HB stars at a given color
in both M3 and M13 have similar O. It seems that one might have to
invoke a second ``second parameter'' beyond helium mixing to explain
the M13/M3 pair---a not particularly attractive idea.

In summary, nothing simple works.

\acknowledgments

We are grateful to Peter Stetson and Don VandenBerg for providing
results prior to publication. RTR \& BD are supported in part by NASA
Long Term Space Astrophysics Grant NAG5-700 and NAGW-4106 and
STScI/NASA Grant GO-5903.  The financial support by the {\it Agenzia
Spaziale Italiana} (ASI) is gratefully acknowledged.


\end{document}